# Digital Nudge Effects in Optimal Stopping Games: Antagonistic Interactions in Layered Choice Architectures

Batisse Manhardt and Ja-Nae Duane

## Abstract

*Digital nudges have become a prominent area of research in human-centered interaction, yet little is known about how they interact when deployed together. This study examines the individual and combined effects of anchoring, salience, and social proof digital nudges on stopping behavior and outcome quality in a sequential decision task. Using a $2^3$ between-subjects factorial design grounded in the secretary problem ($N$ = 181), we find that (1) an anchor nudge significantly increased exploration rather than constraining it ($b$ = 1.207, $p$ = .017; (2) the salience nudge produced the smallest effect of any condition; and (3) combining the anchor and social proof nudges produced the worst outcome quality of any condition ($181.83 vs. $192.21 for anchor-only), with outcome quality declining monotonically as nudge count increased. These findings provide the first controlled evidence for a digital nudge crowding-out effect with direct implications for the ethical design of digital interfaces.*

## 1. Introduction

When a consumer searches for an item on Amazon, they encounter an engineered decision environment. For example, an item may appear to be marked as “Amazon’s Choice” alongside a counter showing only five of the items left in stock. No option is removed, yet the consumer’s decision has already been guided before they consciously engage with it.

This is digital nudging in practice. Digital nudges combine data, information, interactions, or incentives with design elements to leverage cognitive biases (Duane et al., 2024). A subset of digital nudges is dark patterns, characterized by their intent to guide users and to act against users' best interests (Duane et al., 2024; Gray et al., 2023). In 2024, the $2.5 billion FTC settlement against Amazon brought these tactics into regulatory focus, finding that the company used deceptive tactics to lead consumers to unintentionally purchase Amazon Prime subscriptions while making cancellation exceedingly difficult (Federal Trade Commission, 2025). Yet the underlying mechanics of these digital nudges and their subcategory of dark patterns, how they interact, amplify, or undermine one another, remain poorly understood.

Digital nudges and dark patterns are now pervasive in all environments. Whether it is a fitness app or a supermarket’s website, nearly all corporations are aware of and actively utilize these. As Duane et al. (2024) argue, digital nudges are best understood as instances in which data, information, interactions, or incentives are combined with design elements within a digital environment to leverage cognitive biases or heuristics to influence a choice. Caraban et al. (2019) identify numerous variables that affect nudging design and effectiveness, further demonstrating the nuanced nature of the digital nudging space (Schneider et al., 2018; Weinmann et al., 2016).

Despite this, nudges have largely been studied in isolation, examining single nudge types within single contexts, leaving open the question of how nudges interact when deployed together. In practice, consumers rarely encounter a single nudge on its own (Duane, 2023). Instead, they encounter layered combinations of defaults, social proof signals, anchors, and loss-framed cues. Whether these combinations amplify, neutralize, or antagonize one another remains an open empirical question. This paper addresses that gap with a controlled experiment grounded in the secretary problem (Gilbert & Mosteller, 1966), a well-validated sequential search-stop paradigm that isolates stopping behavior and outcome quality across varying choice architectures. Using a 23 between-subjects factorial design ($N$ = 181), we tested three digital nudges (anchoring, salience, and social proof) and their seven possible combinations, comparing each against a no-nudge control. Our findings have direct implications for the future of human-computer interaction research and ethical interface design. As regulatory scrutiny following the Federal Trade Commission (Federal Trade

Commission, 2025) makes clear, understanding how combinations of nudges work and fail is no longer merely an academic question.

Search-stop studies have previously been conducted, but their main objective has been to test ideas such as exploration-exploitation trade-offs, in which optimal stopping theory is explored (Bouhlel, 2022; Goldstein et al., 2019; Rapoport et al., 2022). Such research has evolved to examine various environments, including competitive ones (Rios & Ghosh, 2024), as well as how heuristics influence individuals' decisions (Gurney et al., 2025). However, although this research examines the cognitive domain of stop-search behavior, there is no literature on the effectiveness of digital nudges in these environments. Digital nudges occur when data, information, interactions, or incentives are combined with design elements in a digital environment to leverage cognitive biases or heuristics to influence user choices (Duane et al., 2024). This design logic can be pervasive and may utilize dark patterns that work against users' interests (Gray, 2023; Schneider et al., 2018) through the same cognitive mechanisms identified in digital nudge research: exploiting defaults, social proof, and loss aversion to drive spending and engagement.

One of the most common digital nudges is the use of defaults. Previous studies have examined the effectiveness of salience in relation to cookie-setting preferences on websites (Laine, 2021), whether visual cues can lead to early quitting in visual searches (Moher et al., 2024), and status quo bias in decision-making (Krieger & Felder, 2013). While several of the above studies may have implications for the digital nudging domain, they do not effectively examine a user's interactions with a salience heuristic in a search-stop environment; rather, they focus on a specific use case, whether that be cookie-setting acceptance or health insurance policies. Caplin and Martin (2012) found that the default effect is strong when individuals are inattentive but erodes when they become attentive. Similarly, others have found that the strength of the default effect diminishes significantly as the duration increases (White et al., 2021). Salience nudges operate by directing System 1 attention toward a preselected option (Kahneman, 2011), though their effectiveness diminishes when users are attentive and engaged (Caplin & Martin, 2012; White et al., 2021), suggesting that motivated players may be less susceptible than passive users.

Anchoring nudges leverage users' tendency to over-rely on salient reference information when making sequential judgments (Mochon & Frederick, 2013; Tversky & Kahneman, 1974). Tversky and Kahneman (1974) first introduced the concept, studying how selecting a random number affected an individual's response to a question they may not be knowledgeable about. More recent literature has examined individuals' overall search effort (Gurney et al., 2025) and anchoring in sequential decision-making (Mochon & Frederick, 2013). Specifically, Mochon and Frederick (2013) measured how judgments serve as anchors for subsequent judgments, not how anchoring through demonstration may affect an individual's search-stop actions.

Social proof uses consensus information to guide behavior under uncertainty (Cialdini, 2007) and is especially potent when the quality of options is unknown. This is similar to the condition that sequential search creates (Venema et al., 2020). When combined with loss-framed anchoring, social uncertainty should amplify sensitivity to negative framing, producing larger deviations from optimal stopping than either nudge alone. Finally, while individual nudge effects are reasonably documented, combination effects are not. Howley and Ocean (2021) found non-linear interactions between nudge types, and Hummel and Maedche (2019) identified crowding as a mechanism by which additional nudges reduce effectiveness rather than compounding it. In digital interfaces that routinely layer multiple nudge types simultaneously, cognitive overload may cause combinations to antagonize rather than reinforce one another (Schneider et al., 2018). Based on the literature above, we propose the following hypotheses for this study:

**H1:** The salience nudge will produce a smaller shift in stopping behavior and outcome quality than anchoring or social proof nudges, as attentive participants in a high engagement task are less susceptible to default-style cues (Caplin & Martin, 2012; White et al., 2021).

**H2:** The anchor nudge will cause participants to stop earlier in the sequence, as visible progress anchors operationalize loss aversion by making skipped options increasingly salient (Mochon & Frederick, 2013; Tversky & Kahneman, 1974).

**H3:** The combination of social proof and loss-framed anchoring will produce greater deviation from the control than either nudge alone, as social uncertainty amplifies sensitivity to loss-framed information (Howley & Ocean, 2021; Venema et al., 2020).

**H4:** Combining multiple nudges will not yield additive improvements in outcome quality; instead, combinations will exhibit antagonistic effects as the number of nudges increases, reflecting cognitive overload in interactive decision environments (Hummel & Maedche, 2019; Weinmann et al., 2016).

## 3. Study Design

A between-subjects design crossed three binary nudges: an anchor nudge, a social nudge, and a salience nudge, yielding eight experimental conditions for this game. The game was conducted through Qualtrics, and participants were randomly assigned to one condition and one of three box sets (stimulus sets containing ten monetary values each). All three box sets were counterbalanced across conditions.

### 3.1. Participants and Recruitment

Participants were recruited via an anonymous Qualtrics survey link through a convenience sample. Eligibility required participants to be between 18 and 25 years old, targeting young adults. No compensation was provided for this pilot of the game. This study had 267 participants. Of 267 participants, 47 did not complete the game. Thirty-three did not pass the attention check, where they were asked how many total boxes there were (10), whether they could return to a previously selected box (No), and what the objective of the game was (select the box with the highest monetary value). Finally, 6 participants were excluded for outlier time to complete the survey (greater than 900 seconds). The final sample for this study was 181.

### 3.2. Variables

Two primary dependent variables were used to capture participants' behavior. The first was a stopping box (accepted box), which was the box number at which the participant accepted a value (range 1–10). This variable operationalizes stopping behavior relative to the search sequence. The second dependent variable was the accepted amount ($), representing the dollar value of the box the participant accepted (range $160 to $201). This operationalizes outcome quality. Finally, we had a robustness measure, the within-set percentile rank, which was constructed by ranking each participant's accepted amount within their box set (on a 0–1 scale). This controls for stimulus-difficulty confounds more completely than box-set dummy variables.

## 4. Results

With a bimodal coefficient of 0.624, the distribution is strongly bimodal, exceeding the 0.555 threshold.

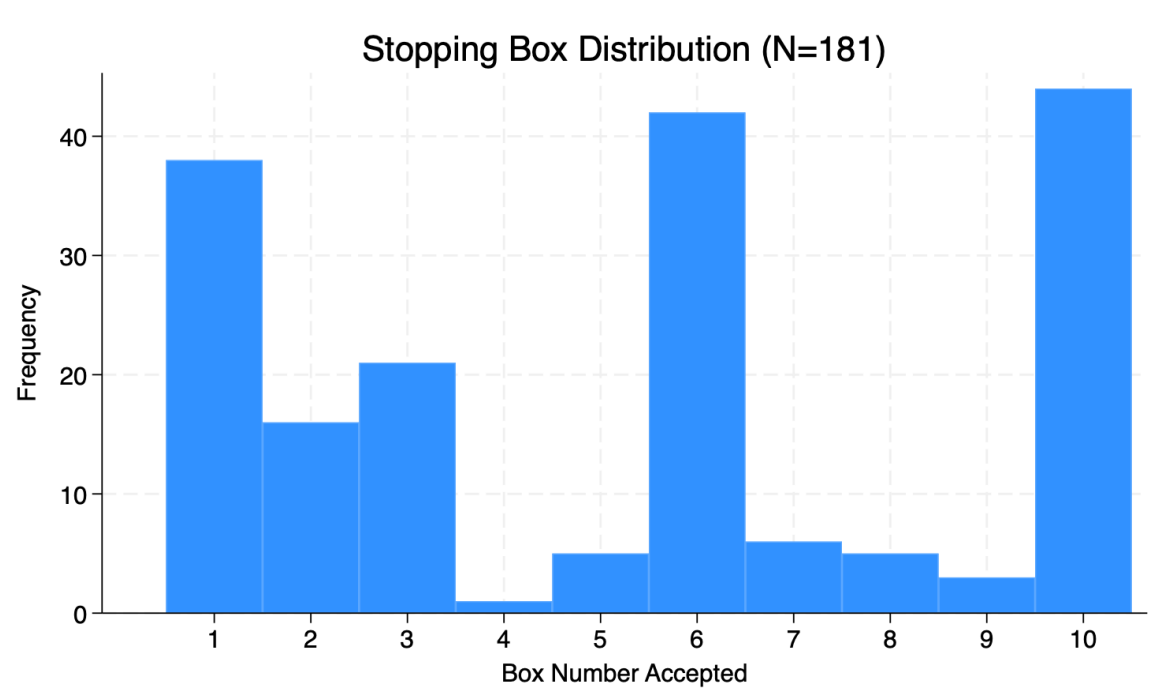

**Figure 1. Stopping box distribution**

Two patterns arise: 41.4% of participants stopped at boxes 1–3 (very early termination), and 50.3% stopped at boxes 6, 9, or 10 (late termination). Notably, only 3.3% stopped in the "optimal" range derived from the secretary problem. The dip between the two peaks of the distribution is consistent with many participants being unfamiliar with the secretary problem. From these two patterns, two potential strategies may explain the distribution: stopping quite early or having a higher risk tolerance and waiting until the absolute end, referred to as "Satisficers" and "Exhaustive searchers" respectively.

Risk tolerance had no significant interaction with any nudge, despite a borderline main effect on stopping behavior ($r$ = +0.15, $p$ = 0.040), indicating that higher risk tolerance is associated with stopping later in the sequence. However, there is no clear relationship between risk tolerance and the accepted box, further indicating that nudge effects are independent of one's risk tolerance towards the accepted box number.

### 4.1. Main Effects: The Power of Anchoring

Anchoring was found to significantly increase the average stopping box by 1.3, reversing the hypothesized decrease in average stopping box effect.

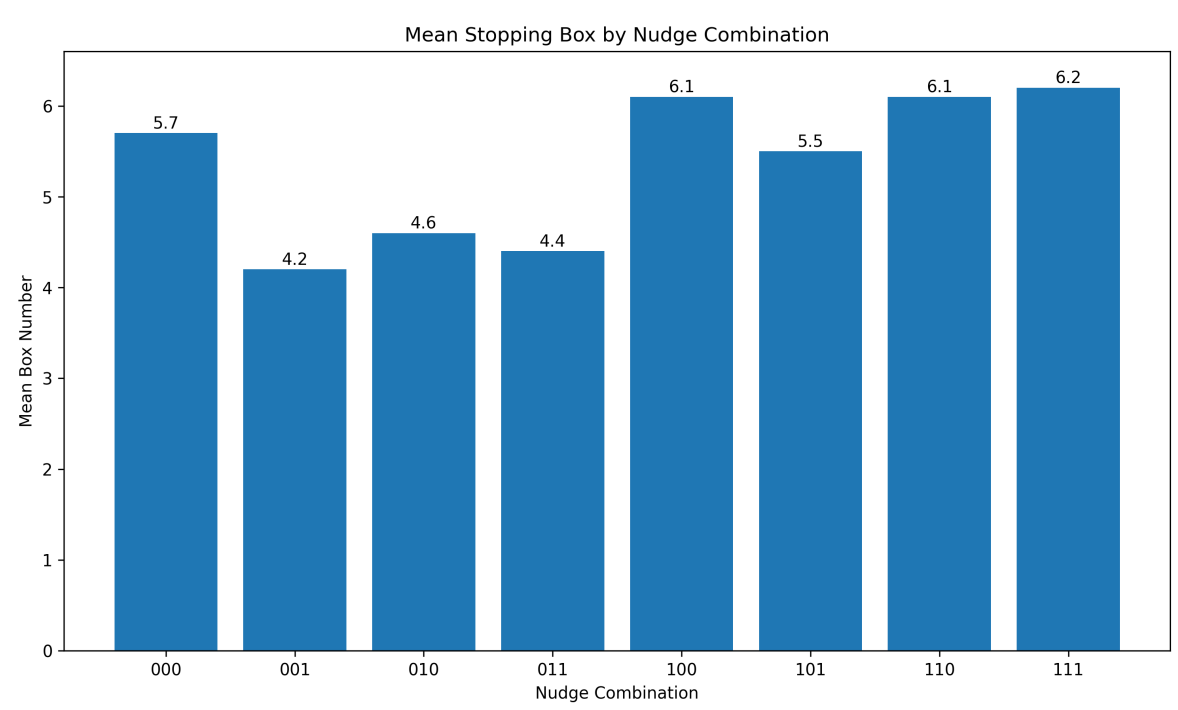

**Figure 2. Mean stopping box by nudge combination**

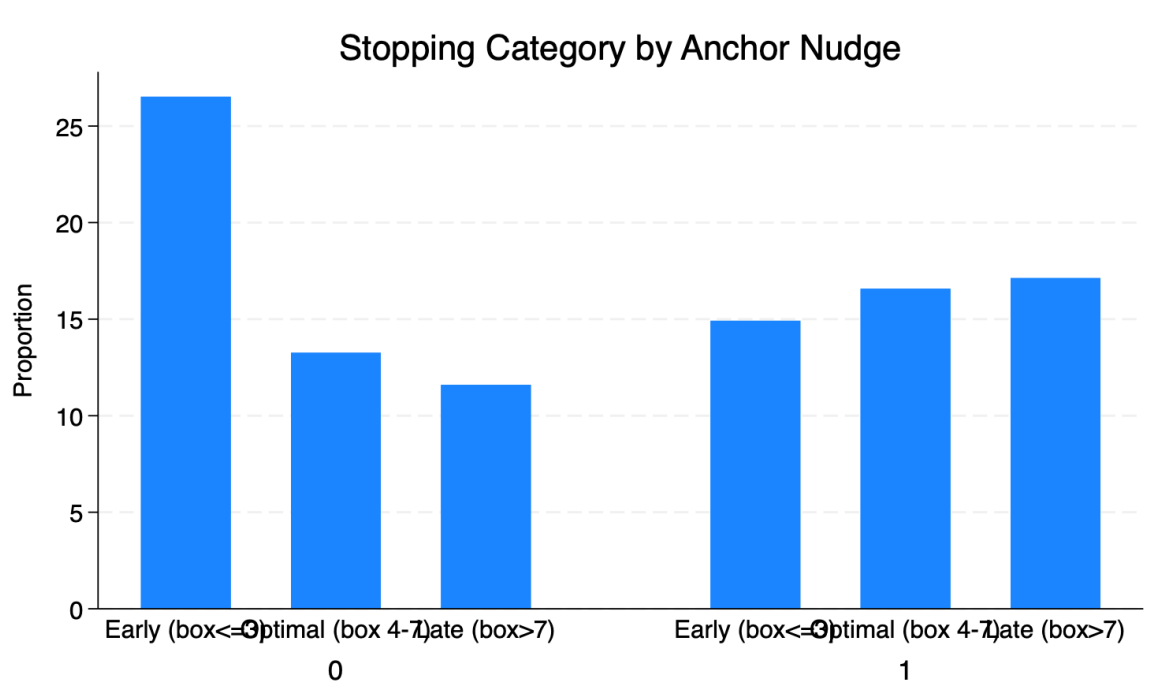

**Figure 3. Stopping category by anchor nudge**

The anchor nudge significantly increased the stopping box by approximately 1.3 boxes on average ($t$ = 2.39, $p$ = .018, $d$ = 0.36). This effect was confirmed non-parametrically (Mann-Whitney $U$ = 3,326, $p$ = .027, $r_{rb}$ = 0.19). The finding indicates that the anchor nudge served as an exploration tool rather than its intended effect, which was to encourage a sense of loss aversion. Individuals may have been encouraged to continue opening boxes after seeing that they had only opened a few out of the total number.

### 4.2. Interaction Effects: Antagonistic Nudges

The anchor-only condition (100) produced the highest mean accepted amount in the study ($192.21; $d$ = +0.67) compared with the control. However, the anchor-social proof condition (110) interestingly reversed this effect: the mean accepted amount fell to $181.83, the lowest of any condition ($d$ = 0.33 vs. control). Utilizing within-set percentile rank DV, condition 110 yields the lowest mean rank of any condition (0.366, vs. 0.600 for anchor-only and 0.500 for control), and the anchor × social interaction term is statistically significant ($\beta$ = 0.256, $p$ = .048). H3 is partially supported by a significant anchor-x-social-proof interaction, but the interaction is antagonistic rather than amplifying, ultimately reversing the expected direction.

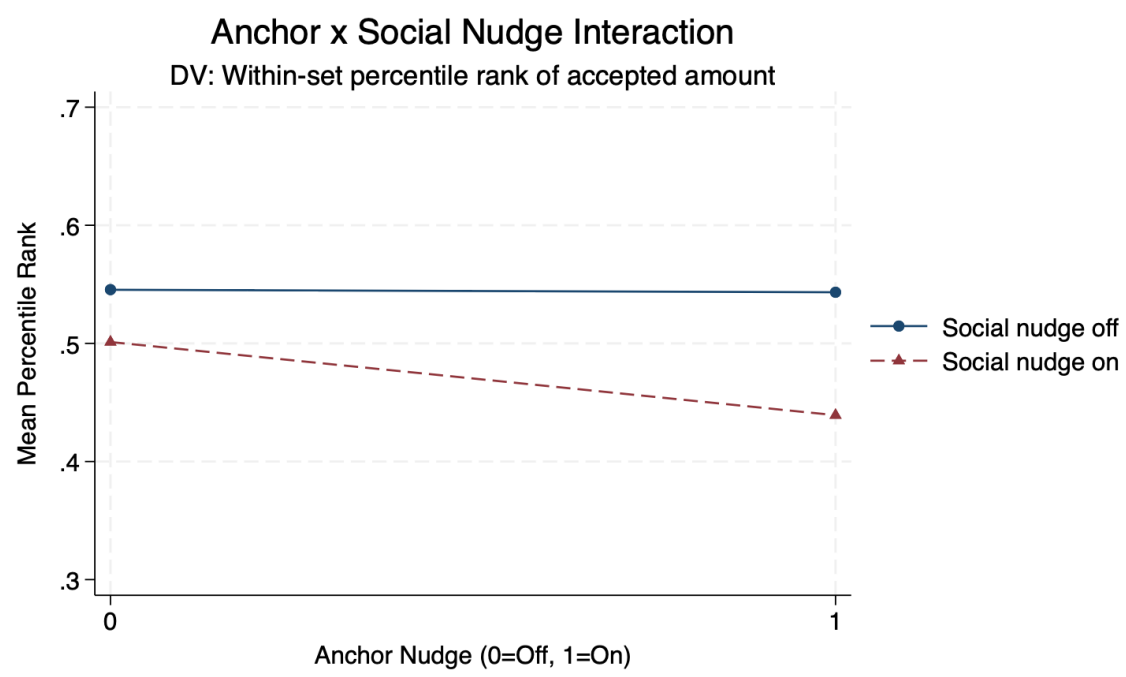

**Figure 4. Anchor x social nudge interaction**

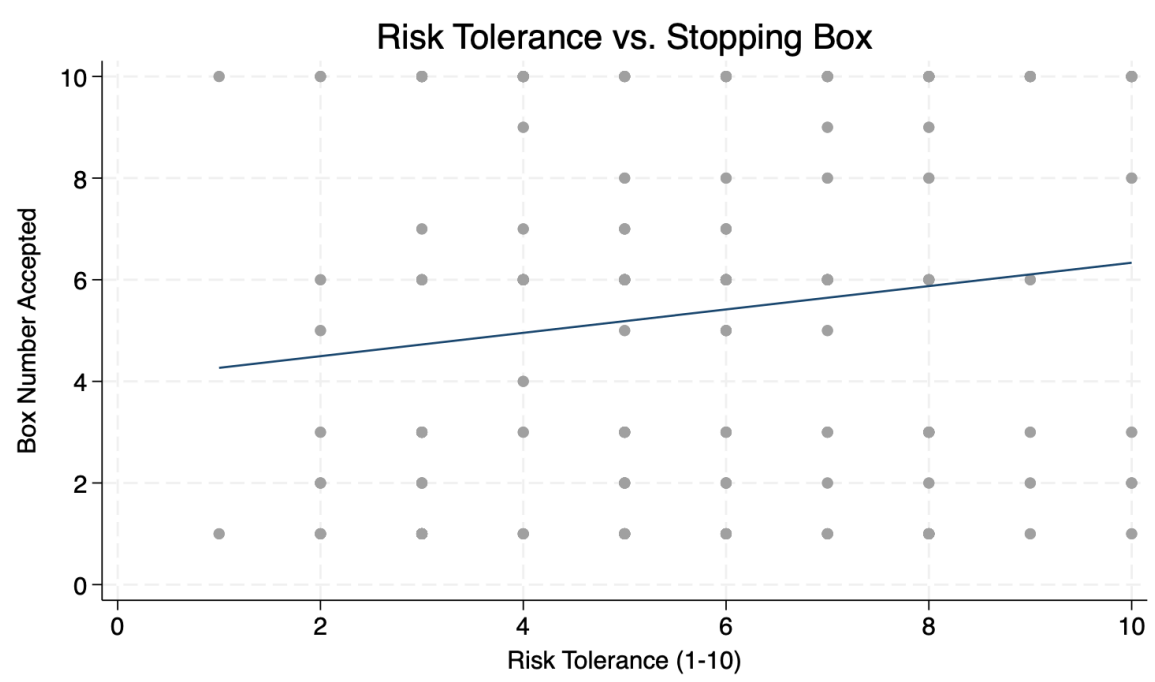

**Figure 5. Dose-Response: mean stopping box by nudge count**

H4 is supported on outcome quality: the accepted amount peaks at one nudge ($188.91) and declines monotonically with additional nudges, consistent with a potential crowding out effect. However, H4 is not supported for stopping behavior. The anchor + social combination (condition 110) does not differ significantly from anchor-alone (100) or social-alone (010) on the stopping box, only outcome quality. Dose-response analysis shows that while a single nudge reduces the average stopping box, two or three nudges have the opposite effect, leading to later stopping. Stopping box shows a non-monotonic relationship with nudge count (OLS $\beta = 0.27$, $p = .351$; Spearman $r = .059$, $p = .432$).

## 4.3. Independence of Strategy and Quality

**Table 1. Mediation analysis: anchor nudge on outcome quality**

| Path | Analysis Step | Estimate ($b$) | S.E. | $p$-value |
|---|---|---|---|---|
| $a$-path | Anchor → Stopping Box | 1.207 | 0.499 | .017* |
| $b$-path | Stopping Box → Amount | −0.107 | 0.244 | .655 |
| $c$-path | Total Effect (Anchor → Amount) | −1.167 | 2.471 | .466 |
| $c'$-path | Direct Effect (Controlled) | −1.038 | 2.397 | .529 |
| Product | Indirect Effect ($a \times b$) | −0.129 | — | — |

*Note.* $N = 181$. $*p < .05$. The non-significant $b$-path confirms that search strategy (stopping box) and outcome quality (amount) are statistically independent.

Using Baron and Kenny's (1986) four-step approach, we find that while the anchor nudge significantly predicts the stopped box ($b = 1.207$, $p = 0.017$), the stopping box does not significantly predict the accepted box amount, indicating independence between search strategy and outcome quality.

The primary hypotheses are tested using Ordinary Least Squares (OLS) regressions. Table 2 presents the estimated effects of digital nudges on stopping behavior and outcome quality, controlling for individual risk tolerance and box-set heterogeneity. Robustness checks using Ordered Probit specifications were conducted to account for the non-normality of the stopping data.

Model 1 assesses the factors that determine the exact sequence number at which a player terminates their search. In this fully saturated interaction model, the direct coefficients for the main treatments, Anchor Nudge ($\beta = 0.589$, SE = 0.994), Social Proof Nudge ($\beta = -0.928$, SE = 1.038), and Salience Nudge ($\beta = -1.256$, SE = 0.942), do not reach traditional levels of statistical significance. Furthermore, the two-way interaction terms (Anchor × Social = 0.805; Anchor × Salience = 0.428; Social × Salience = 1.335) and the three-way interaction term (−0.281) fail to show a reliable shift in the baseline box stoppage number ($p > 0.10$).

Model 2 explores the experimental effects on the monetary value ($) of the accepted box, achieving notably higher explanatory power ($R^2 = 0.177$). While the standalone main treatments for the Anchor, Social Proof, and Salience nudges show no statistically significant baseline direct effects on the accepted value ($p > 0.10$), dynamics emerge when these treatments are combined. Specifically, the regression in Model 2 reveals an antagonistic relationship when the anchor and social proof mechanisms are combined. The interaction coefficient for Anchor × Social is negative and statistically significant ($\beta = -8.631$, SE = 4.592, $p < 0.10$), indicating that exposing participants to both nudges concurrently reduces the final accepted amount relative to what would be predicted by their independent effects. Interestingly, this effect is partially reversed when all three nudges are deployed together; the three-way interaction term (Anchor × Social × Salience) exhibits a positive, statistically significant coefficient ($\beta = 11.090$, SE = 6.525, $p < 0.10$). The baseline constant for outcome quality is highly significant at 191.960 ($p < 0.01$) and closely aligns with the distribution's mean properties after partialing out demographic and fixed-effect controls.

**Table 2. Main Regression Results: Sensitivity Analysis ($N = 181$)**

| | (1) Box Number | (2) Accepted Amount |
|---|---|---|
| *Main Treatments* | | |
| Anchor Nudge | 0.589 | 4.065 |
| | (0.994) | (3.232) |
| Social Proof Nudge | −0.928 | −0.114 |
| | (1.038) | (3.232) |
| Salience Nudge | −1.256 | 2.800 |
| | (0.942) | (3.145) |
| *Interaction Effects* | | |
| Anchor × Social | 0.805 | −8.631* |
| | (1.530) | (4.592) |
| Anchor × Salience | 0.428 | −7.215 |
| | (1.457) | (4.676) |
| Social × Salience | 1.335 | −3.318 |
| | (1.419) | (4.749) |
| Three-Way Interaction | −0.281 | 11.090* |
| | (2.120) | (6.525) |
| Demographic Controls[a] | Yes | Yes |
| Box-Set Fixed Effects[b] | Yes | Yes |
| Constant | 4.299** | 191.960*** |
| | (1.889) | (6.423) |
| Observations | 181 | 181 |
| $R^2$ | 0.068 | 0.177 |

Robust standard errors in parentheses. $*p < 0.10$, $**p < 0.05$, $***p < 0.01$. $a$ Includes controls for Gender, Risk Tolerance, and Secretary Problem Familiarity.
$b$ Controlled for stimulus set heterogeneity (Box Sets 1–3).
*Note.* Six observations were excluded due to implausible durations (>900 seconds).

Overall, these OLS specifications point to a subtle choice-architecture environment in which digital nudges do not predominantly operate as isolated, additive forces but rather manifest through pathways that alter the quality of outcomes.

## 5. Discussion

This study set out to examine how digital nudges, individually and in combination, shape stopping behavior and outcome quality in a sequential decision environment. Three findings stand out. The most striking result is the anchor reversal. The box counter nudge, designed to operationalize loss aversion by making skipped options increasingly salient, instead significantly increased exploration ($b$ = 1.207, $p$ = .017). We attribute this to the Goal-Gradient Effect (Kivetz et al., 2006): rather than perceiving opened boxes as foregone opportunities, players appear to have read the counter as progress toward a goal, motivating continued search. This has direct implications for interface design. Progress indicators ubiquitous in e-commerce environments, such as rewards trackers or bundling options, may systematically encourage exploration rather than commitment, depending on how they are framed. Critically, reversing the anchor's direction, showing remaining rather than opened boxes, may produce the loss-framing effect originally hypothesized and represents a clear target for future work.

The salience nudge produced the smallest effect of any condition, consistent with H1 and with prior evidence that default-style cues lose power when users are attentive and engaged (Caplin & Martin, 2012; White et al., 2021). The search-stop task, which allowed deliberate System 2 thinking without time pressure, appears to have reduced the System 1 susceptibility that salience nudges rely on. The antagonistic interaction between the anchor and social proof nudges is perhaps the most consequential finding for practice. Anchor-only participants achieved the highest outcome quality in the study ($192.21, $d$ = +0.67 vs. control), while the anchor-social proof combination produced the lowest ($181.83, $d$ = 0.33 vs. control). The dose-response analysis reinforces this: outcome quality peaked with one active nudge and declined monotonically with each additional nudge. This crowding-out pattern, consistent with predictions of cognitive overload (Hummel & Maedche, 2019; Schneider et al., 2018), suggests that layering nudges may actively harm individual outcomes rather than compound positive effects.

However, the three-way interaction term (Anchor × Social × Salience)'s partial reversal of this antagonistic effect created by the anchor-social proof combination suggests that through careful design, potential harm to player outcomes may be mitigated.

## 6. Conclusion and Future Work

There are several limitations to this study. While this study establishes a framework for investigating digital nudges and their antagonistic interactions, several limitations offer opportunities for future research within digital environments. First, this study had large effect sizes. Subsequent studies should utilize larger sample sizes to adequately powered comparisons and verify our findings. Second, digital environments should more deeply operationalize core behavioral constructs such as anchoring, scarcity, and loss aversion, thereby solidifying the theoretical link between digital nudging and the cognitive systems that drive user decision-making.

This study investigated the effects of digital nudges on individuals in a sequential search-stop game environment. We find that a box counter anchor nudge significantly increases the number of boxes opened, overshadowing a weaker salience default nudge and, conversely, producing the worst outcome when combined with a social proof nudge. We find that, consistent with prior research, System 2 cues have the potential to override System 1 cues through digital nudging but have the potential to elicit overcrowding effects. By providing a rare direct comparison of competing digital nudges within a shared framework, this work establishes a critical foundation for future research on the boundary conditions and interaction effects of behavioral heuristics in digital environments.

## 7. References

Baron, R. M., & Kenny, D. A. (1986). The moderator–mediator variable distinction in social psychological research: Conceptual, strategic, and statistical considerations. Journal of Personality and Social Psychology, 51(6), 1173–1182. https://doi.org/10.1037/0022-3514.51.6.1173

Bouhlel, I., Chessa, M., Festré, A., & Guerci, E. (2022). When to stop searching in a highly uncertain world? A theoretical and experimental investigation of "two-way" sequential search tasks. Journal of Economic Behavior & Organization, 203, 80–92. https://doi.org/10.1016/j.jebo.2022.08.034

Caplin, A., & Martin, D. (2012). Defaults and attention: The drop out effect (NBER Working Paper No. 18534). National Bureau of Economic Research. https://www.nber.org/papers/w18534

Caraban, A., Karapanos, E., Gonçalves, D., & Campos, P. (2019). 23 ways to nudge: A review of technology-mediated nudging in human-computer interaction. In Proceedings of the 2019 CHI Conference on Human Factors in Computing Systems (pp. 1–15). https://doi.org/10.1145/3290605.3300733

Cialdini, R. B. (2007). Descriptive social norms as underappreciated sources of social control. Psychometrika, 72(2), 263–268. https://doi.org/10.1007/s11336-006-1560-6

Duane, J. (2023). Digital nudges: An investigation of both consumer and designer perspectives (Doctoral dissertation). ProQuest Dissertations Publishing. https://search.proquest.com/openview/b0fef013420febb8f880c43dc7e1b91e

Duane, J., Ericson, J., & McHugh, P. (2024). Digital nudges: A systematic narrative review and taxonomy. Behaviour & Information Technology. https://doi.org/10.1080/0144929X.2024.2440116

Federal Trade Commission. (2025). FTC secures historic $2.5 billion settlement against Amazon. https://www.ftc.gov/news-events/news/press-releases

Gilbert, J. P., & Mosteller, F. (1966). Recognizing the maximum of a sequence. Journal of the American Statistical Association, 61(313), 35–73. https://doi.org/10.2307/2283044

Goldstein, D. G., McAfee, R. P., Suri, S., & Wright, J. R. (2019). Learning when to stop searching. Management Science, 66(3), 1124–1138. https://doi.org/10.1287/mnsc.2018.3245

Gray, C. M., Sánchez Chamorro, L., Obi, I., & Duane, J.-N. (2023). Mapping the landscape of dark patterns scholarship: A systematic literature review. In Proceedings of the Designing Interactive Systems Conference. https://doi.org/10.1145/3563703.3596635

Gurney, N., Miller, J. H., & Pynadath, D. V. (2025). Exploring the choice landscape: Anchoring and framing effects on search behavior in complex choices. Journal of Choice Modelling, 55, 100549. https://doi.org/10.1016/j.jocm.2025.100549

Howley, P., & Ocean, N. (2021). Can nudging only get you so far? Testing for nudge combination effects. European Review of Agricultural Economics, 49(5), 1113–1139. https://doi.org/10.1093/erae/jbab041

Hummel, D., & Maedche, A. (2019). How effective is nudging? A quantitative review on the effect sizes and limits of empirical nudging studies. Journal of Behavioral and Experimental Economics, 80, 47–58. https://doi.org/10.1016/j.socec.2019.03.005

Kahneman, D. (2011). Thinking, fast and slow. Farrar, Straus and Giroux.

Kivetz, R., Urminsky, O., & Zheng, Y. (2006). The goal-gradient hypothesis resurrected: Purchase acceleration, illusionary goal progress, and customer retention. Journal of Marketing Research, 43(1), 39–58. https://doi.org/10.1509/jmkr.43.1.39

Krieger, M., & Felder, S. (2013). Can decision biases improve insurance outcomes? An experiment on status quo bias in health insurance choice. International Journal of Environmental Research and Public Health, 10(6), 2560–2577. https://doi.org/10.3390/ijerph10062560

Laine, J. (2021). There is no decision: Design of cookie consent banner and its effect on user consent (Master's thesis, Tampere University). Tampere University Institutional Repository.

Mochon, D., & Frederick, S. (2013). Anchoring in sequential judgments. Organizational Behavior and Human Decision Processes, 122(1), 69–79. https://doi.org/10.1016/j.obhdp.2013.04.002

Moher, J., Delos Reyes, A., & Drew, T. (2024). Cue relevance drives early quitting in visual search. Cognitive Research: Principles and Implications, 9(1), 42. https://doi.org/10.1186/s41235-024-00587-1

Rapoport, A., Seale, D. A., & Spiliopoulos, L. (2023). Progressive stopping heuristics that excel in individual and competitive sequential search. Theory and Decision, 94(3), 415–439. https://doi.org/10.1007/s11238-022-09881-0

Rios, I., & Ghosh, P. (2024). Competition in optimal stopping: Behavioral insights. Manufacturing & Service Operations Management, 26(6), 2256–2273. https://doi.org/10.1287/msom.2022.0621

Schneider, C., Weinmann, M., & vom Brocke, J. (2018). Digital nudging: Guiding online user choices through interface design. Communications of the ACM, 61(7), 67–73. https://doi.org/10.1145/3198382

Tversky, A., & Kahneman, D. (1974). Judgment under uncertainty: Heuristics and biases. Science, 185(4157), 1124–1131. https://doi.org/10.1126/science.185.4157.1124

Venema, T. A. G., Kroese, F. M., Benjamins, J. S., & de Ridder, D. T. D. (2020). When in doubt, follow the crowd? Responsiveness to social proof nudges in the absence of clear preferences. Frontiers in Psychology, 11, 1385. https://doi.org/10.3389/fpsyg.2020.01385

White, B. X., Jiang, D., & Albarracín, D. (2021). The limits of defaults: The influence of decision time on default effects. Social Cognition, 39(5), 543–569. https://doi.org/10.1521/soco.2021.39.5.543